\documentclass[]{elsart}
\usepackage{amssymb}
\usepackage{pst-node,pstricks,graphicx}

\begin{document}                

\begin{frontmatter}
\title{
An upper limit to the photon fraction in cosmic rays
above $10^{19}$~eV from the Pierre Auger Observatory
}

\author{J.~Abraham$^{6}$,} \author{M.~Aglietta$^{41}$,}
\author{C.~Aguirre$^{8}$,} \author{D.~Allard$^{73}$,}
\author{I.~Allekotte$^{1}$,} \author{P.~Allison$^{69}$,}
\author{C.~Alvarez$^{44}$,} \author{J.~Alvarez-Mu\~{n}iz$^{58}$,}
\author{M.~Ambrosio$^{38}$,} \author{L.~Anchordoqui$^{68,\: 79}$,}
% ...

\begin{abstract}                

An upper limit  of 16\% (at 95\% c.l.) is derived for the photon fraction in 
cosmic rays with energies greater than $10^{19}$~eV, based on observations
of the 
depth of shower maximum performed with the hybrid detector of the Pierre Auger 
Observatory. This is the first such limit on photons obtained by observing the 
fluorescence light profile of air showers. This upper limit confirms and 
improves on previous results from the Haverah Park and AGASA surface arrays.
Additional data recorded with the Auger surface detectors
for a subset of the event sample support the conclusion that a 
photon origin of the observed events is not favored.
\end{abstract}

\end{frontmatter}



\section{Introduction}
\label{sec-intro}

The origin of ultra-high energy (UHE) cosmic rays above $10^{19}$~eV is still 
unknown~\cite{reviews}. Their energy spectrum, arrival directions
and composition can be inferred from air shower observations.
However, 
agreement has not yet been reached on whether there is a break in the energy spectrum 
around $E_{\rm GZK} \sim 6 \times 10^{19}$~eV (= 60 EeV). Such a 
steepening in the energy spectrum  is expected
if UHE cosmic rays come from cosmologically distant sources~\cite{gzk},
as is suggested by 
their overall isotropy. There have been claims, as yet unconfirmed, for clustering
on small angular scales, and correlations with possible classes of sources.
Moreover, results concerning the nuclear composition are still inconclusive.

While this deficit of robust observational results is partly due to the
extremely small fluxes and, correspondingly, small numbers of events at
such high energies, discrepancies might arise also from the different
experimental techniques used.
For instance, the determination of the primary energy from the ground 
array alone relies on the comparison
with air shower simulations and is thus prone
to uncertainties in modelling high energy interactions. Therefore it is essential
to test results from air shower observations independently.
The present work provides just such a cross-check for the upper limit 
derived previously from ground arrays on the photon fraction in UHE cosmic rays.
An upper limit is set on the photon fraction above 10~EeV which is 
twice as strong as those given previously.

Photons are expected to dominate over nucleon primaries in non-acceleration 
(``top-down'') models of UHE cosmic-ray  origin~\cite{bhat-sigl,sarkar03,models}
which have been invoked in particular to account for a continuation
of the flux above $E_{\rm GZK}$ without a spectral feature as indicated by
AGASA data~\cite{agasa-gzk}.
Thus, the determination of the photon 
contribution is a crucial probe of cosmic-ray source models. 
Separating photon-induced showers from events initiated by nuclear 
primaries is experimentally much easier than distinguishing light and
heavy nuclear primaries. As an example, average depths of shower maxima at
10~EeV primary energy are predicted to be
about 1000~g~cm$^{-2}$, 800~g~cm$^{-2}$, and 700~g~cm$^{-2}$ 
for primary photons, protons, and iron nuclei, respectively.
Moreover, analyses of nuclear composition are uncertain due to our poor knowledge
of hadronic interactions at very high energies.
Photon showers, being driven mostly by electromagnetic interactions,
are less affected by such uncertainties and can be modelled with
greater confidence.
To avoid the uncertainty from modelling
hadronic interactions, we adopt an analysis method that does not require the
simulation of nuclear primaries but compares data to photon simulations
only.

So far limits on the UHE photon fraction in cosmic rays
have been set by ground arrays alone.
By comparing the rates of near-vertical showers to inclined ones
recorded by the Haverah Park shower detector, upper limits (95\% c.l.) of
48\% above 10~EeV and 50\% above 40~EeV were deduced~\cite{ave}.
Based on an analysis of muons in air showers observed by the
Akeno Giant Air Shower Array (AGASA), the upper limits (95\% c.l.) to the photon
fraction were estimated to be 28\% above 10~EeV and 67\% above
32~EeV ~\cite{shinozaki}.
An upper limit of 67\% (95\% c.l.) above 125~EeV was derived in a 
dedicated study of the highest energy AGASA events~\cite{risse05}.

In this work, we obtain a photon limit from the {\em direct} observation
of the shower profile with fluorescence telescopes,
using the depth of shower maximum $X_{\rm max}$ as the
discriminating observable.
To achieve a high accuracy in reconstructing the shower geometry,
we make use of the ``hybrid'' detection technique, i.e. we
select events observed by both the ground array and the
fluorescence telescopes~\cite{mostafa}.
For a subset of the event sample, a sufficient number of ground detectors were
also triggered, yielding a variety of additional shower
observables. Considering as example the signal risetime measured with the
ground array, we demonstrate the discrimination power of these independent
observables to photon-induced showers.

The plan of the paper is as follows.
In Section~\ref{sec-photons}, predictions for the UHE photon fraction in cosmic-ray
source models and features of photon-initiated air showers are summarized.
Section~\ref{sec-data} contains the description of the data and
of photon simulations. In particular, the data selection criteria are discussed.
A careful choice of the quality and fiducial volume cuts is required to
control a possible experimental bias for photon primaries.
In Section~\ref{sec-results}, the method for deriving a photon fraction
is described and applied to the data.
An example of the discrimination power of observables registered by
the surface array is shown in Section~\ref{sec-sd}.
Finally in Section~\ref{sec-outlook}, we discuss the prospects for improving the 
bound on UHE photons.


\section{Photons as cosmic-ray primaries}
\label{sec-photons}

\begin{figure}[t]
\begin{center}
\includegraphics[height=13.0cm,angle=270]{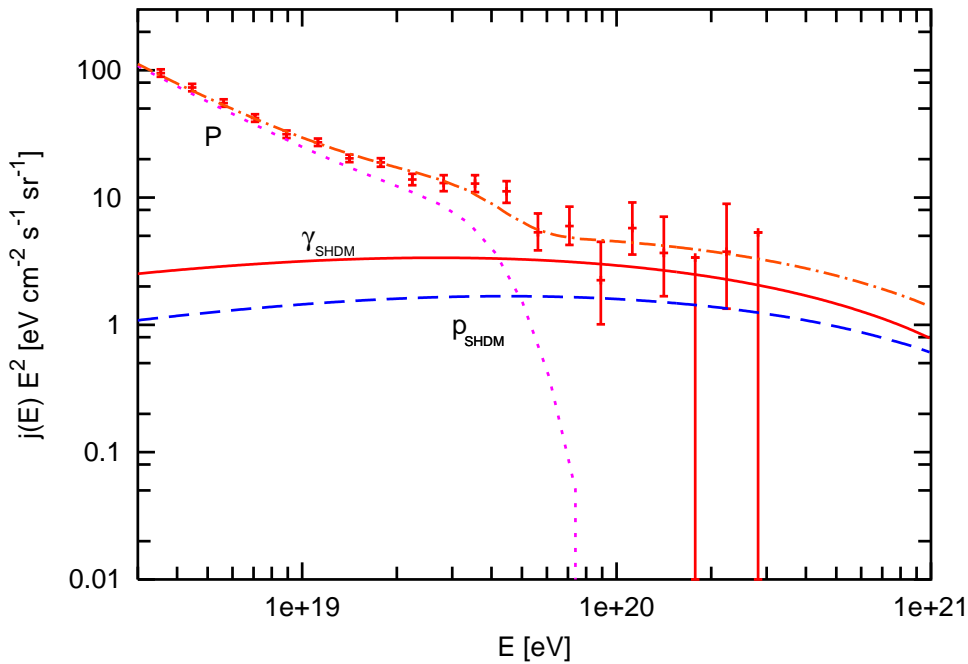}
\caption{
Example of a SHDM model fit to AGASA 
data~\cite{agasa-gzk} (in the highest and third highest energy bins which have zero
events, upper flux limits are shown). The spectra of photons ($\gamma_{_{\rm SHDM}}$)
and protons ($\rm p_{_{\rm SHDM}}$) from SHDM, and
an assumed additional nucleonic component at lower energy (P),
as well as their sum is plotted. 
Photons dominate above $\sim  5 \times 10^{19}$~eV.
(Figure taken from~\cite{models}.)
}
\label{fig-specshdm}
\end{center}
\end{figure}

The theoretical challenge of explaining acceleration of protons
to the highest energies is
circumvented in non-acceleration models~\cite{bhat-sigl}.
A significant fraction of the UHE cosmic rays are predicted by
these models to be photons (see e.g. \cite{sarkar03,models}).
For instance, UHE photons may be produced uniformly in the
universe by the decay/annihilation of relic topological defects (TD)~\cite{td}. 
During propagation to Earth, they interact with background radiation fields and
most of them cascade down to GeV energies where
the extragalactic photon flux is constrained by the EGRET experiment; the 
remaining UHE photons can contribute to the cosmic-ray flux above 10~EeV.
By contrast in the Super Heavy Dark Matter (SHDM) model~\cite{shdm}, the UHE photons 
are generated in the decay of relic metastable particles (such as ``cryptons'' 
\cite{ellis}) which are clustered as cold dark matter in our Galaxy. Since the halo 
is believed to be effectively transparent to such UHE photons, they would be directly
observed at Earth with little processing.
In the Z-Burst (ZB) scenario~\cite{zb}, photons are generated from the resonant 
production of Z bosons by UHE cosmic 
neutrinos annihilating on the relic neutrino background.
A distinctive feature of all
these models is the prediction of a large photon flux at
high energies, as is expected from considerations of QCD fragmentation \cite{frag}. 
As an illustration, Figure~\ref{fig-specshdm} (taken from~\cite{models})
shows a SHDM model fit to the highest energy 
AGASA events; photons are the dominant particle species above
$\sim  5 \times 10^{19}$~eV.

Photons can also be produced in ``conventional'' acceleration models
by the GZK-type process from $\pi^0$ decays.
Typically, the corresponding photon fluxes are relatively small. 
For instance, based on the spectrum obtained by the HiRes
experiment~\cite{hires-gzk}, the expected photon fraction is only of order 
1\% or below~\cite{models}.

It should be noted that the photon flux arriving at Earth for
a specific source model is subject to uncertainties arising from photon
propagation: assumptions concerning the very low frequency (few MHz) radio
background and inter-galactic magnetic fields must be made~\cite{sarkar03,models}. 
The typical range of energy loss lengths usually adopted for photons are
7--15~Mpc at 10~EeV and 5--30~Mpc at 100~EeV.

Ultra-high energy photons can be detected by the particle
cascades they initiate when entering the atmosphere of the Earth.
Compared to air showers initiated by nuclear primaries,
photon showers at energies above 10~EeV
are in general expected to have a larger depth of shower
maximum $X_{\rm max}$ and to contain fewer secondary muons.
The latter is because the mean free paths for
photo-nuclear interactions and direct muon pair production are
more than two orders of magnitude larger than the radiation length. 
Consequently, only a small fraction of the primary energy
in photon showers is generally transferred into secondary hadrons
and muons.

\begin{figure}[t]
\begin{center}
\includegraphics[height=8.7cm,angle=0]{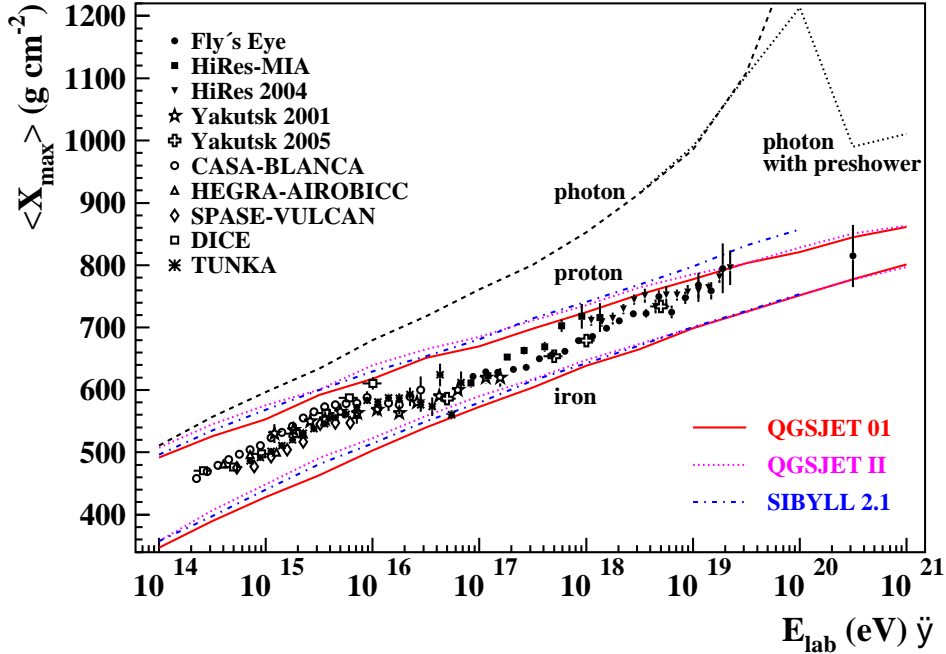}
\caption{
Average depth of shower maximum $<$$X_{\rm max}$$>$ versus energy
simulated for primary photons, protons and iron nuclei.
Depending on the specific particle trajectory through the geomagnetic
field, photons above $\sim  5 \times 10^{19}$~eV can create a
pre-shower: as indicated by the splitting of the
photon line, the average $X_{\rm max}$ values then do not only depend
on primary energy but also arrival direction.
For nuclear
primaries, calculations for different hadronic interaction models
are displayed (QGSJET 01~\cite{qgs01}, QGSJET II~\cite{qgs2}, 
SIBYLL 2.1~\cite{sib21}).
Also shown are experimental data (for references to the experiments, 
see~\cite{xmax-heck}).
}
\label{fig-xmaxvse}
\end{center}
\end{figure}

In Figure~\ref{fig-xmaxvse}, simulated $X_{\rm max}$ values for
showers initiated by primary photons, protons and iron nuclei
are shown as a function of the primary energy.
The large $X_{\rm max}$ values for photon showers at 10~EeV are
essentially due to the small
multiplicity in electromagnetic interactions, in contrast to the large
number of secondaries produced in inelastic interactions of
high-energy hadrons. 
Secondly, because of the LPM effect~\cite{lpm},
the development of photon showers is even further delayed above
$\sim$~10~EeV. 
Another feature of the LPM effect is an
increase of shower fluctuations: $X_{\rm max}$ fluctuations for
photon showers are $\sim$~80~g~cm$^{-2}$ at 10~EeV,
compared to $\sim$~60~g~cm$^{-2}$ and $\sim$~20~g~cm$^{-2}$
for primary protons and iron nuclei, respectively.

At higher energies, cosmic-ray photons may convert in the geomagnetic 
field and create a pre-shower before entering the atmosphere~\cite{erber}.
The energy threshold for geomagnetic conversion is $\sim$~50~EeV
for the Auger southern site.
Conversion probability and pre-shower features depend both on
primary energy and arrival direction.
In the case of a pre-shower, the subsequent air shower
is initiated as a superposition of lower-energy secondary photons and
electrons. For air showers from converted photons, the $X_{\rm max}$
values and the fluctuations are considerably smaller than from single photons 
of same total energy. From the point of view of air shower development, the 
LPM effect and pre-shower formation compete with each other.

In this work, cascading of photons in the geo\-magnetic 
field is simulated with the PRE\-SHO\-WER code~\cite{homola} and
shower development in air, including the LPM effect~\cite{lpm}, is
calculated with CORSIKA~\cite{heck}.
For photo-nuclear processes, we assume the extrapolation of
the cross-section as given by the Particle Data Group~\cite{pdg}, and we employed
QGSJET~01~\cite{qgs01} as a hadron event generator.

\section{The Data Set}
\label{sec-data}

The Auger data used in this analysis were taken with a total
of 12 fluorescence telescopes situated at
two different sites~\cite{bellido}, during the period January 2004
to February 2006. The number of surface detector
stations deployed~\cite{bertou05} grew during this period 
from about 150 to 950.
A detailed description of the Auger Observatory is given 
in~\cite{auger}.

For the present analysis, we selected hybrid events,
i.e. showers observed both with (one or more) surface tanks
and telescopes.
Even when only one tank is triggered, the angular accuracy improves from
$\ge 2^\circ$ for observation with one telescope alone
to $\sim$~0.6$^\circ$ for hybrid detection~\cite{mostafa,bonifazi},
thus reducing significantly the corresponding uncertainty
in the reconstruction of $X_{\rm max}$.

The reconstruction of the shower profiles~\cite{bellido,argiro}
is based on an end-to-end calibration of the
fluorescence telescopes~\cite{brack}.
Monthly models for the atmospheric density profiles are used which
were derived from local radio soundings~\cite{keilhauer}.
An average aerosol model is adopted based on measurements of the
local atmospheric aerosol content~\cite{roberts}.
Cloud information is provided by IR monitors, positioned at
the telescope stations~\cite{roberts}.
Cross-checks on clouds are obtained from
measurements with LIDAR systems (near the telescopes) and with
a laser facility near the center of the array~\cite{roberts,malek}.
The Cherenkov light contribution of the shower is calculated
according to~\cite{nerling}.
An energy deposit profile is
reconstructed for each event. 
A Gaisser-Hillas function~\cite{gh} is fitted to the
profile to obtain the depth of shower maximum, and
the calorimetric shower energy is obtained by integration.
It has been checked that this function provides a reasonable
description
of the simulated shower profiles independent of the primary particle,
provided all four parameters of the Gaisser-Hillas fit
are allowed to vary.

A correction for missing energy, the ``invisible'' energy fraction
carried by neutrinos and high-energy muons, has to be applied.
The fraction of missing energy depends on the primary particle type.
In case of nuclear primaries, the correction amounts to 7--14\%, 
with a slight dependence on primary energy and the
hadronic interaction model used~\cite{barbosa,pierog}.
For photon primaries, the missing energy fraction is much smaller
and amounts to $\sim$~1\%~\cite{pierog}.
We applied the correction assuming photon primaries, so that the
energy threshold chosen in the analysis corresponds to the effective
energy of primary photons.

For the current analysis, the differences between the energy 
estimates for different primaries are relatively small ($\sim$~10\%) due to the
near-calorimetric measurement of the primary energy by the
fluorescence technique.
Moreover, relative to photon showers, the energies of nuclear primaries are
slightly underestimated.
This would slightly {\em deplete} an event sample from showers
ascribed to nuclear primaries or,
correspondingly, increase the number ascribed to photons. Thus, the limit 
derived here for photons is conservative with respect to the missing energy 
correction.
It seems worthwhile to mention that for ground array studies, where
the energy of photons can be underestimated by more than
30\% (see, for instance,~\cite{shinozaki}), such corrections to the
primary energy which depend on the unknown primary particle type must
be treated with particular caution.

The following quality cuts are applied for event selection
(in Appendix~\ref{app-cut}, distributions of cut variables
are displayed):

\begin{itemize}
\item[$\bullet$] 
Quality of hybrid geometry: distance of closest approach of the
reconstructed shower axis to the array tank with the largest signal
$<$1.5~km, and difference between the reconstructed shower front
arrival time at this tank and the measured tank time $<$300~ns;
\item[$\bullet$] Primary energy E$>$10$^{19}$~eV;
\item[$\bullet$] $X_{\rm max}$ observed;
\item[$\bullet$] Number of phototubes in the fluorescence detector
                 triggered by shower $\ge$6;
\item[$\bullet$] Quality of Gaisser-Hillas (GH) profile fit:
    $\chi^2$(GH) per degree of freedom $<$6, and
    $\chi^2$(GH)/$\chi^2$(line)$<$0.9, where $\chi^2$(line) refers
    to a straight line fit;
\item[$\bullet$] Minimum viewing angle of shower direction
                 towards the telescope $>$15$^\circ$;
\item[$\bullet$] Cloud monitors confirm no disturbance of
                 event observation by clouds.
\end{itemize}

\begin{figure}[t]
\begin{center}
\includegraphics[height=6.3cm,angle=0]{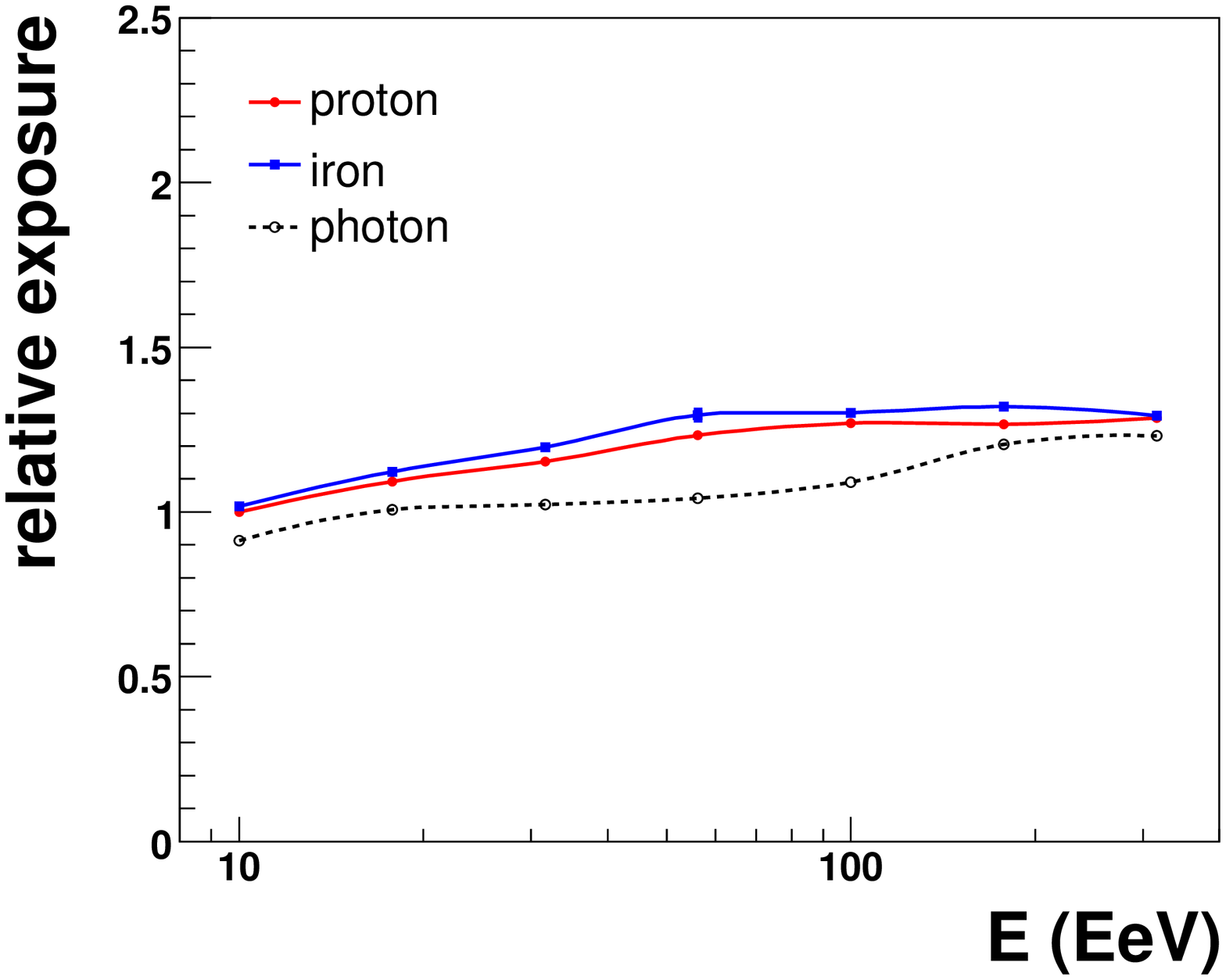}
\includegraphics[height=6.3cm,angle=0]{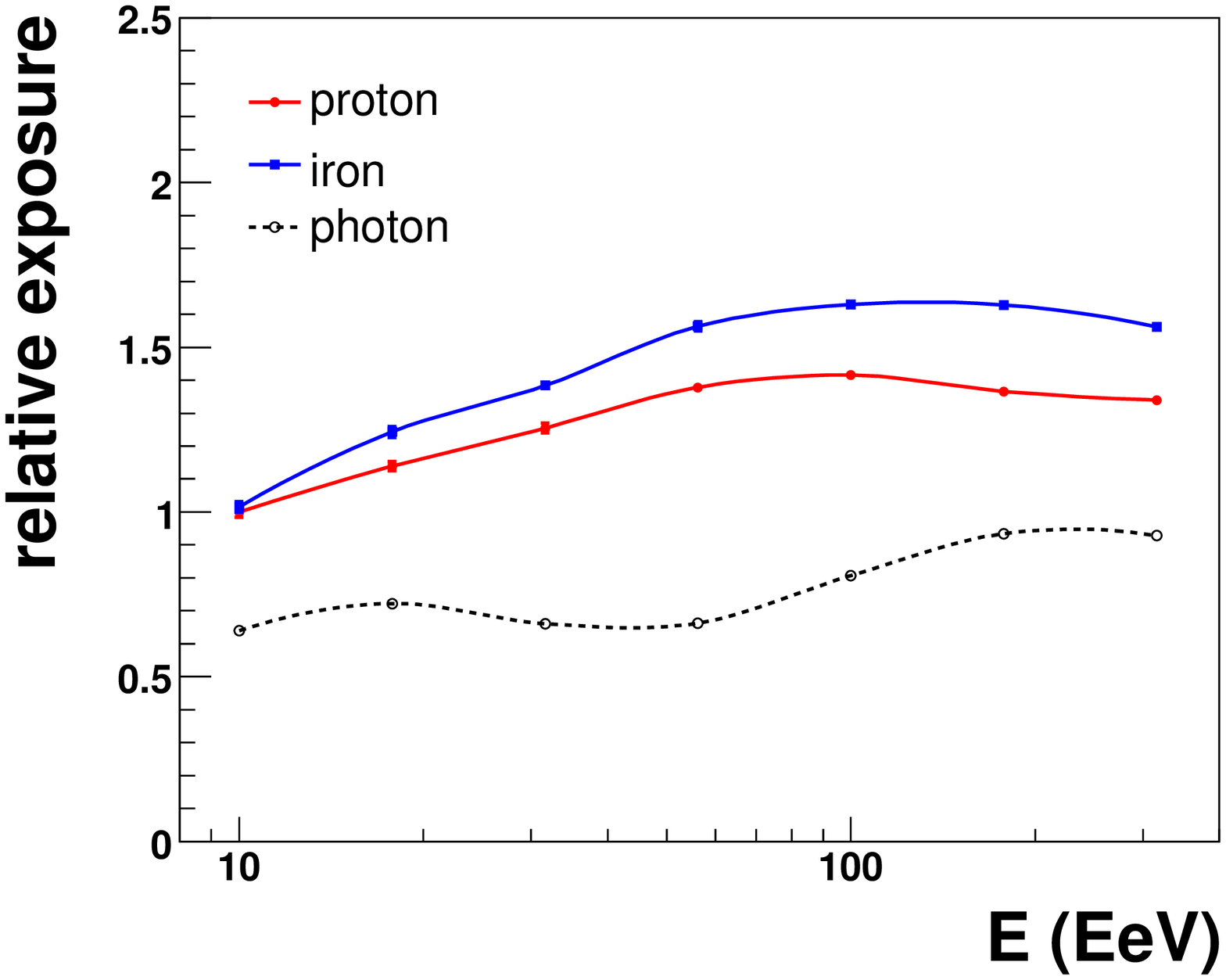}
\includegraphics[height=6.3cm,angle=0]{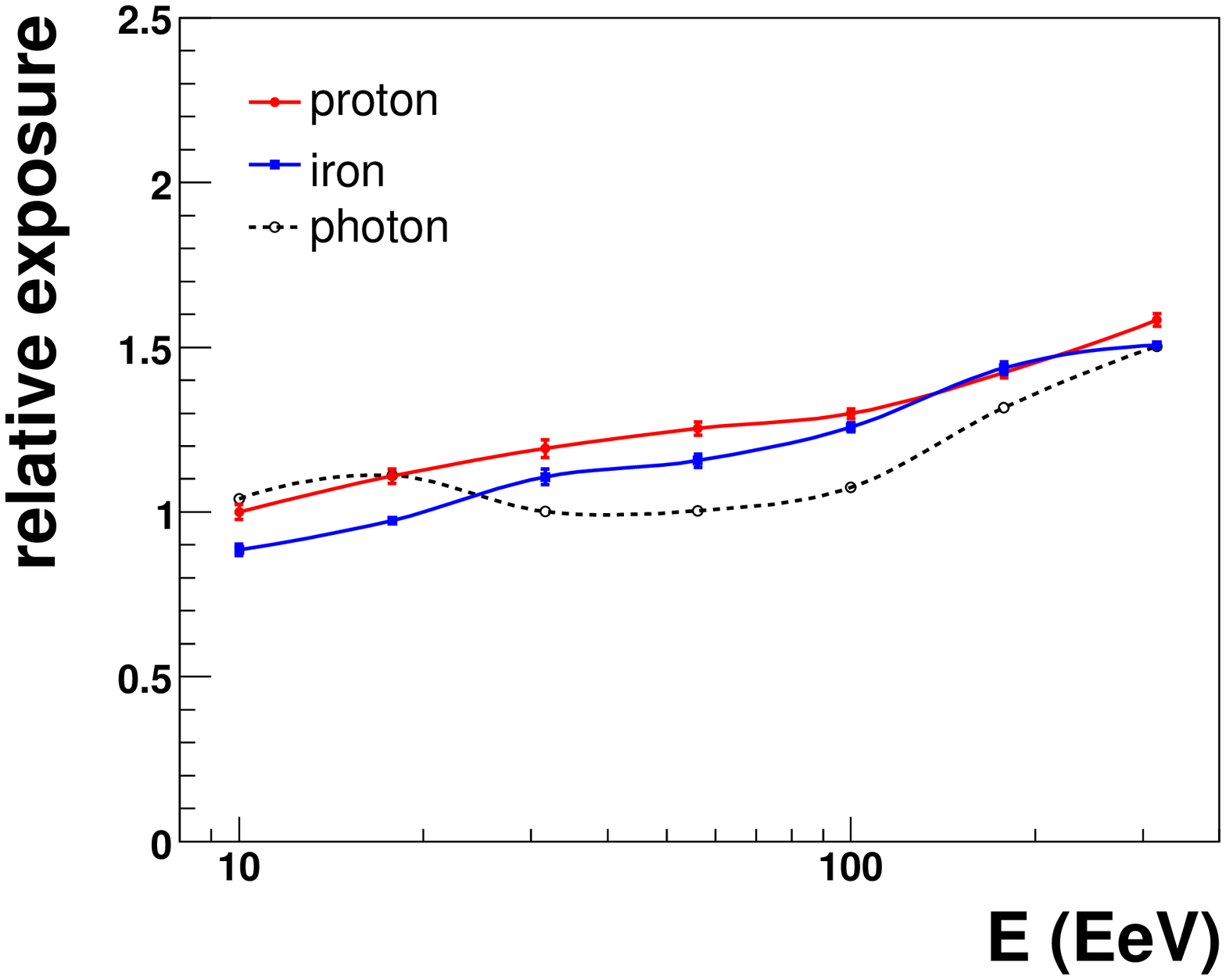}
\caption{
Relative exposures for photon, proton, and iron primaries
as a function of energy after trigger (top), after quality cuts (middle)
and after fiducial volume cuts are applied (bottom) to reduce the bias
against photons.
A reference value of one is adopted for proton at 10~EeV.
}
\label{fig-acc}
\end{center}
\end{figure}

\begin{figure}[t]
\begin{center}
\includegraphics[height=8.7cm,angle=0]{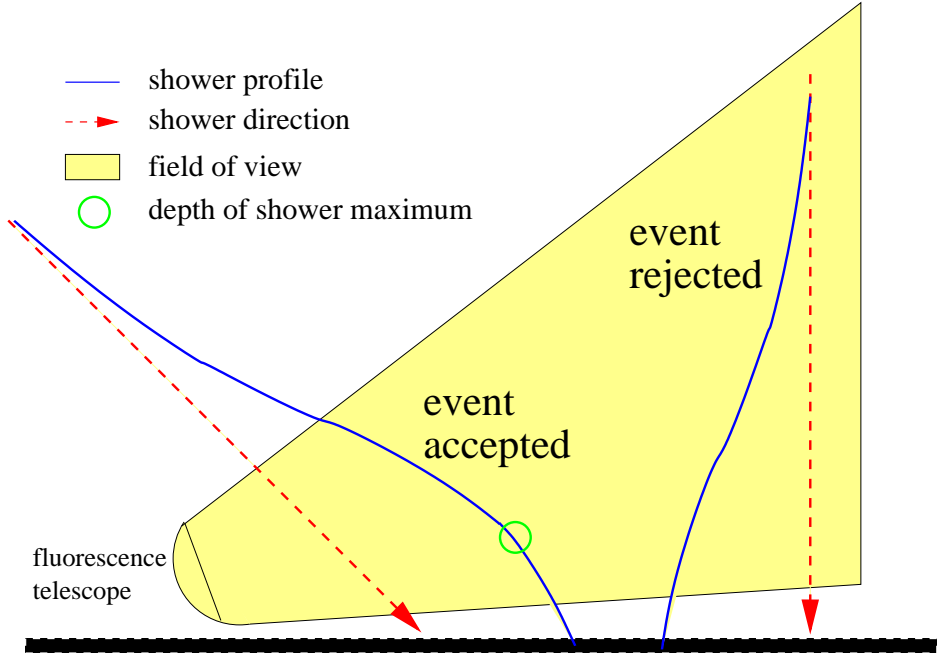}
\caption{
Photon showers and the selection requirement of observing
$X_{\rm max}$.
For near-vertical photon showers, $X_{\rm max}$ is below the
field of view of the telescopes; possibly the showers even reach
ground before being fully developed as in the example shown.
Such photon showers were
rejected by the quality cuts. The situation changes when regarding
more inclined photon events. The slant atmospheric depth that
corresponds to the lower edge of the field of view increases with
zenith. $X_{\rm max}$ can then be reached within the field of view,
and the photon showers pass the $X_{\rm max}$ quality cut.
Requiring a minimum zenith angle in the analysis, the reconstruction
bias for photons is strongly reduced.
}
\label{fig-sketch}
\end{center}
\end{figure}

Care must be taken about a possible bias against photon primaries
of the detector acceptance.
In Figure~\ref{fig-acc} we show the acceptance for photons and
nuclear primaries at different steps of the analysis, computed
using shower simulations 
with the CONEX code~\cite{conex} which reproduces well the CORSIKA
predictions for shower profiles. 
Light emission and propagation through the atmosphere and
the detector response were simulated according to~\cite{prado}.
As can be seen from the Figure, the acceptances are comparable for
all types of primaries after trigger (top plot).
However, after profile quality cuts
(middle plot) the detection efficiency for photons is smaller by
a factor $\sim$2 than for nuclear primaries, because primary photons reach
shower maximum at such large depths (of about 1000~g~cm$^{-2}$,
see Figure~\ref{fig-xmaxvse}) that
for a large fraction of showers the maximum is outside the field of
view of the telescopes.
This holds, in particular, for near-vertical photon showers:
since the Auger Observatory is located at an average
atmospheric depth of 880~g~cm$^{-2}$ (measured at a point close to 
the centre of the array) near-vertical photon showers reach the
ground before being fully developed.
Such photon showers are rejected by the quality cuts,
while most of the showers generated by nuclear primaries (with their smaller 
$X_{\rm max}$) are accepted.
An illustration of the effect of this cut on photon showers is
given in Figure~\ref{fig-sketch}.
To reduce the corresponding bias against photons, near-vertical events
are excluded in the current analysis.
Since the average depth of shower maximum increases with
photon energy before the onset of pre-shower,
a mild dependence of the minimum zenith angle with energy
is chosen (see below).

For similar reasons, a cut on distant events is
introduced. The telescopes do not observe shower portions near the
horizon, as the field of view is
elevated by $\sim$~1.5$^\circ$. Thus, the atmospheric depth
which corresponds to the lower edge of the field of view
of a telescope decreases with distance.
Another source of a bias against photon showers is due to
fluorescence light absorption.
The brightest parts of the shower profile, i.e.~those around shower
maximum, are for photon showers
generally closer to the ground. The line of sight towards the shower
maximum traverses regions of higher air density. Hence, for similar
geometrical distances to the shower maximum, the light signal
of the deeper photon showers is more attenuated than for
nuclear primaries.
The consequence is that the distance range below which the telescopes
are fully efficient for detecting showers of a given energy, is
smaller for photon primaries than for nuclear primaries.
This range increases with primary energy.
Thus, an energy-dependent distance cut is applied for the data
selection, in addition to excluding showers at small zenith angles:

\begin{itemize}
\item[$\bullet$] Zenith angle
                 ~$>$35$^\circ~+~g_1(E)$, with
                 $g_1(E)= 10(\lg E/$eV$-19.0)^\circ$
                 for $\lg E/$eV$\le19.7$ and
                 $g_1(E)=7^\circ$ for $\lg E/$eV$>$19.7;
\item[$\bullet$] Maximum distance of telescope to shower impact point
                 ~$<$24~km~+~$g_2(E)$, with
                 $g_2(E)= 12(\lg E/$eV$-19.0)$~km.
\end{itemize}

The acceptances after the fiducial volume cuts are applied are shown in
Figure~\ref{fig-acc} (bottom plot).
The differences between photons and nuclear primaries are now
significantly reduced, with the acceptances being comparable
at energies 10--20~EeV.
With increasing energy, the acceptance for nuclear primaries shows
a modest growth, while the photon acceptance is quite flat in
the investigated energy range. 
Only a minor dependence on the nuclear particle type (proton or
iron) is seen.
Comparing photons to nuclear primaries, the minimum ratio of 
acceptances is $\epsilon_{\rm min} \simeq 0.80$ at energies 50--60~EeV.
At even higher energies, the pre-shower
effect becomes increasingly important, and acceptances for
photons and nuclear primaries become more similar.

The acceptance curves shown in Figure~\ref{fig-acc} can be used
to correct for the detector acceptance when comparing a
measured photon limit to model predictions,
using the model energy spectra as an input.
Since the acceptance ratios after the fiducial volume cuts are not
far from unity, and since the photon acceptance is quite flat in the
energy range
below 100~EeV, the corresponding corrections are expected to be relatively
small and to differ very little between typical model predictions.
In this work, to obtain an experimental limit to the photon fraction
without relying on assumptions on energy spectra of different primaries,
a correction to the photon limit is applied
by conservatively adopting the minimum ratio of acceptances
$\epsilon_{\rm min}$ (a detailed derivation of the approach
is given in Appendix~\ref{app-acc}).

\begin{figure}[t]
\begin{center}
\includegraphics[height=8.7cm,angle=0]{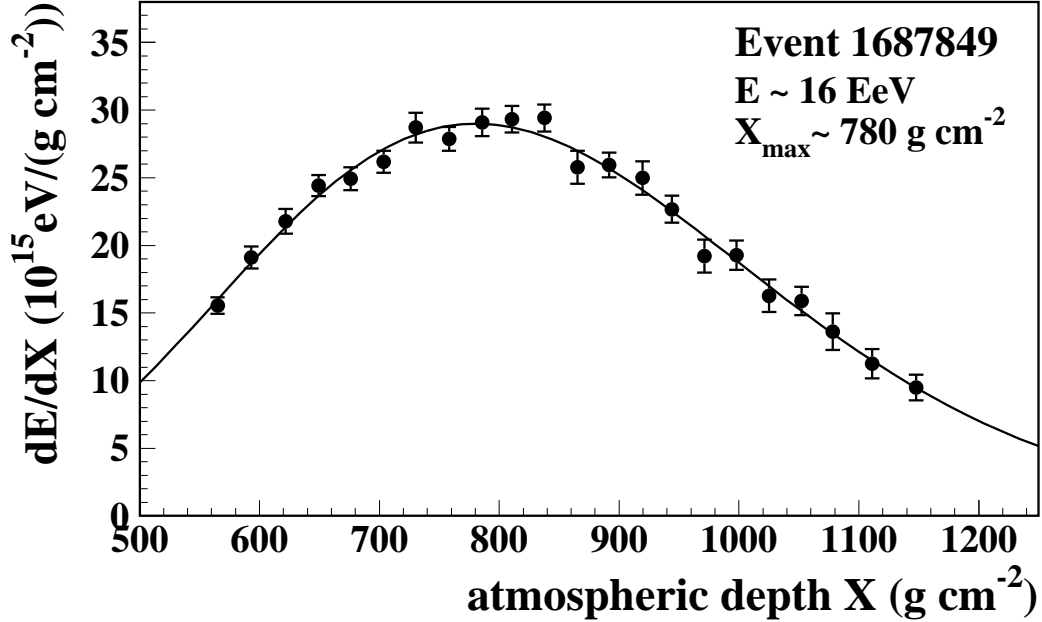}
\caption{
Example of a reconstructed longitudinal energy deposit profile
(points) and the fit by a Gaisser-Hillas function (line).
}
\label{fig-evprof}
\end{center}
\end{figure}

\begin{figure}[t]
\begin{center}
\includegraphics[height=8.7cm,angle=0]{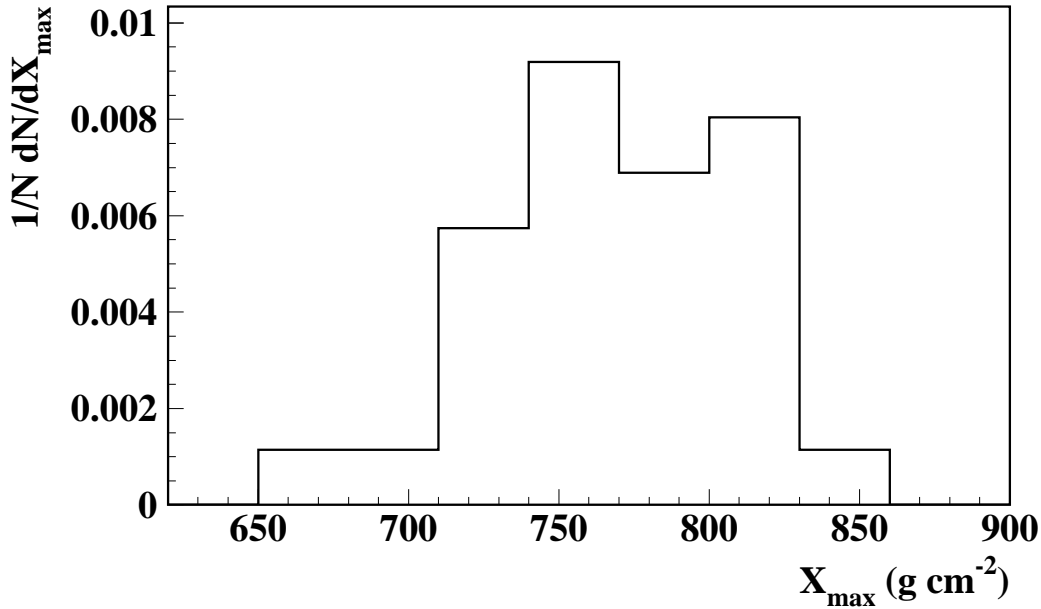}
\caption{
Distribution of $X_{\rm max}$ values of the 29 selected events.
}
\label{fig-xmaxdata}
\end{center}
\end{figure}

Applying the cuts to the data, 29 events with energies
greater than 10~EeV satisfy the selection criteria.
Due to the steep cosmic-ray
spectrum, many events in the sample do not exceed 20 EeV.
The main shower characteristics are summarised for all events in
Table~\ref{tab-data}.
Figure~\ref{fig-evprof} shows the longitudinal profile of an event
reconstructed with 16~EeV and $X_{\rm max} = 780$~g~cm$^{-2}$.
The $X_{\rm max}$ distribution of the selected events is displayed
in Figure~\ref{fig-xmaxdata}.

\begin{table}[t]
{\scriptsize
\begin{center}
\caption{
Event identifier, primary energy, and depth of shower maximum
$X_{\rm max}$ for the selected events.
Also given are the mean depth of shower maximum $<$$X_{\rm max}^{\gamma}$$>$
and its rms fluctuation $\Delta X_{\rm max}^{\gamma}$ predicted from
simulations assuming primary photons. 
In the last column, the differences $\Delta_\gamma$ (in standard
deviations) between photon prediction and data are listed (see text).
A caveat is given in the text concerning the use of these data for
elongation rate studies.
}
\label{tab-data}
\vskip 0.5 cm
\begin{tabular}{cccccc}
\hline
Event ID & Energy  & $X_{\rm max}$  &
 $<$$X_{\rm max}^{\gamma}$$>$ & $\Delta X_{\rm max}^{\gamma}$ & $\Delta_\gamma$
\\
 & [x$10^{18}$~eV] & [g~cm$^{-2}$] & [g~cm$^{-2}$] & [g~cm$^{-2}$]
 & [std.~dev.]
\\
\hline\hline
 668949& 17& 765&    985&  71 & 2.9
\\
 673409& 12& 760&    996&  82 & 2.7
\\
 705583& 11& 678&    973&  77 & 3.6
\\
 737165&202& 821&    948&  27 & 3.3
\\
 828057& 13& 805&    978&  68 & 2.4
\\
 829526& 12& 727&    996&  85 & 3.0
\\
 850018& 54& 774&   1050& 120 & 2.2
\\
 931431& 24& 723&   1022&  89 & 3.2
\\
 935108& 14& 717&    992&  68 & 3.8
\\
 986990& 15& 810&   1000&  87 & 2.1
\\
1109855& 16& 819&   1019&  95 & 2.0
\\
1171225& 15& 786&    993&  74 & 2.6
\\
1175036& 17& 780&   1001& 100 & 2.1
\\
1257649& 10& 711&    971&  76 & 3.2
\\
1303077& 13& 709&    992&  85 & 3.1
\\
1337921& 18& 744&   1029&  93 & 2.9
\\
1421093& 25& 831&   1028&  93 & 2.0
\\
1535139& 15& 768&    998&  77 & 2.8
\\
1539432& 12& 787&    975&  76 & 2.3
\\
1671524& 13& 806&    978&  77 & 2.1
\\
1683620& 20& 824&   1035&  80 & 2.5
\\
1683856& 18& 763&    981&  92 & 2.3
\\
1684651& 12& 753&    991&  79 & 2.8
\\
1687849& 16& 780&   1001&  71 & 2.9
\\
1736288& 10& 726&    981&  71 & 3.3
\\
1826386& 17& 747&    994&  84 & 2.8
\\
1978675& 10& 740&    978&  76 & 2.9
\\
2035613& 11& 802&    998&  90 & 2.1
\\
2036381& 27& 782&   1057& 101 & 2.6
\\
\hline
\\
\end{tabular}
\end{center}
}
\end{table}

For the conditions of the highest-energy event in the sample, event 737165
(see also~\cite{matthews}) with a reconstructed energy of 202~EeV assuming
primary photons, the probability of photon conversion in the geomagnetic field
is $\sim$~100\%. Consequently, the simulated value of the average
depth of shower maximum is relatively small, and shower fluctuations are
considerably reduced.

It should be noted that the event list given in Table~\ref{tab-data}
results from selection criteria optimized for the current primary photon
analysis.  These data cannot be used for studies such as elongation rate 
measurements without properly accounting for acceptance biases.
For instance, the minimum zenith angle required in this
analysis depletes the data sample from showers with
relatively small depths of shower maximum, with the effect being
dependent on primary energy.

The uncertainty $\Delta X_{\rm max}$ of the reconstructed depth of
shower maximum is composed of several contributions, some of which
may vary from event to event.
In this work, we adopt conservative, overall estimates for the current
statistical and systematic uncertainties which
are applied to all selected events.
These uncertainties are expected to decrease significantly in the future.
However, even when adopting conservative estimates,
the present analysis is not limited by the measurement
uncertainties but by event statistics.
This is due to the fact that shower fluctuations for photons are
considerably larger than the measurement uncertainties.

Main contributions to $\Delta X_{\rm max}$ are the uncertainties
in the profile fit, in shower geometry and in
atmospheric conditions (see Table~\ref{tab-xmaxunc}).
Uncertainties in the $X_{\rm max}$ reconstruction from atmospheric conditions
arise from using average models of the density profiles (monthly
averages) and of the aerosol content.  The effect on $X_{\rm max}$
is studied by changing the
atmospheric models and repeating the event reconstruction. The
statistical uncertainty in the determination of the {\it average} model
results in a systematic uncertainty of the $X_{\rm max}$ reconstruction;
it amounts to $\sim 8$~g~cm$^{-2}$ ($\sim 3$~g~cm$^{-2}$ from density
profiles, $\sim 7$~g~cm$^{-2}$ from aerosol model).
A larger uncertainty comes from the {\it spread} around the
averages due to time variations of atmospheric conditions (a detailed
discussion of the density profile variations can be found 
in~\cite{keilhauer}). This results in a statistical uncertainty
of the reconstructed $X_{\rm max}$ value of $\sim 12$~g~cm$^{-2}$
($\sim 6$~g~cm$^{-2}$ from density profiles, $\sim 10$~g~cm$^{-2}$
from aerosol model).

An uncertainty in the $X_{\rm max}^{\gamma}$ values 
predicted from photon simulations results from the uncertainty
in the reconstructed primary energy.
Currently, the systematic uncertainty in energy is 25\%~\cite{bellido}.
For an elongation rate of $\sim$~130~g~cm$^{-2}$ per energy
decade for photons above 10~EeV, this corresponds
to a systematic uncertainty of $\sim$~13~g~cm$^{-2}$.
The elongation rate for primary photons (see Figure~\ref{fig-xmaxvse})
is relatively large
here due to the LPM effect. At highest energies, the elongation rate
decreases with the onset of photon pre-shower in the geomagnetic
field.

It should be noted that this contribution to the systematic
uncertainty from the energy reconstruction
does not refer to the observed $X_{\rm max}$ value
itself. Rather, it enters indirectly in the analysis
since the primary energy is needed as simulation input.

Another uncertainty comes from the extrapolation of
the photo-nuclear cross-section to high energy.
Larger values than adopted here for the cross-section would make showers
initiated by photons more similar to nuclear primaries and
reduce the predicted values for $X_{\rm max}^{\gamma}$.
Based on recent theoretical work on the maximum possible rise of the
photo-nuclear cross-section with energy~\cite{strikman}
an uncertainty of $\sim$~10~g~cm$^{-2}$ is estimated for the predicted depths of 
shower maximum for photons~\cite{rissec2cr}.

Contrary to the case of nuclear primaries, uncertainties from
modelling  high-energy hadron interactions
are much less important in primary photon showers.
From simulations using different hadron event generators,
an uncertainty of $\sim$~5~g~cm$^{-2}$ is adopted.

Adding in quadrature the individual contributions
(see Table~\ref{tab-xmaxunc}) gives a statistical uncertainty 
$\Delta X_{\rm max}^{\rm stat}\simeq$ 28~g~cm$^{-2}$
and a systematic uncertainty
$\Delta X_{\rm max}^{\rm syst}\simeq$ 23~g~cm$^{-2}$.

\begin{table}[t]
\begin{center}
\caption{
Conservative estimates of the contributions to the statistical and
systematic uncertainty of depth of shower maximum for the data and
for the photon simulations.
}
\label{tab-xmaxunc}
\vskip 0.5 cm
\begin{tabular}{lcc}
Data  & $\Delta X_{\rm max}^{\rm stat}$ [g~cm$^{-2}$]
      & $\Delta X_{\rm max}^{\rm syst}$ [g~cm$^{-2}$]
\\
\hline
Profile fit & 20 & 10
\\
Atmosphere & 12 &  8
\\
Geometry reconstruction & 10 & 5
\\
Others          & 10 & 5 
\\
\\
Simulation & &
\\
\hline
Reconstructed energy of event & 5 & 13
\\
Photo-nuclear cross-section & - & 10
\\
Hadron generator            & - & 5
\\
\hline
Total & 28 & 23
\\
\end{tabular}
\end{center}
\end{table}

For each event, 100 showers were simulated as photon primaries.
Since photon shower features can depend in a non-trivial way
on arrival direction and energy, the specific event conditions
were adopted for each event.
Results of the photon simulations are also listed in
Table~\ref{tab-data}.


\section{Results}
\label{sec-results}

\begin{figure}[t]
\begin{center}
\includegraphics[height=8.7cm,angle=0]{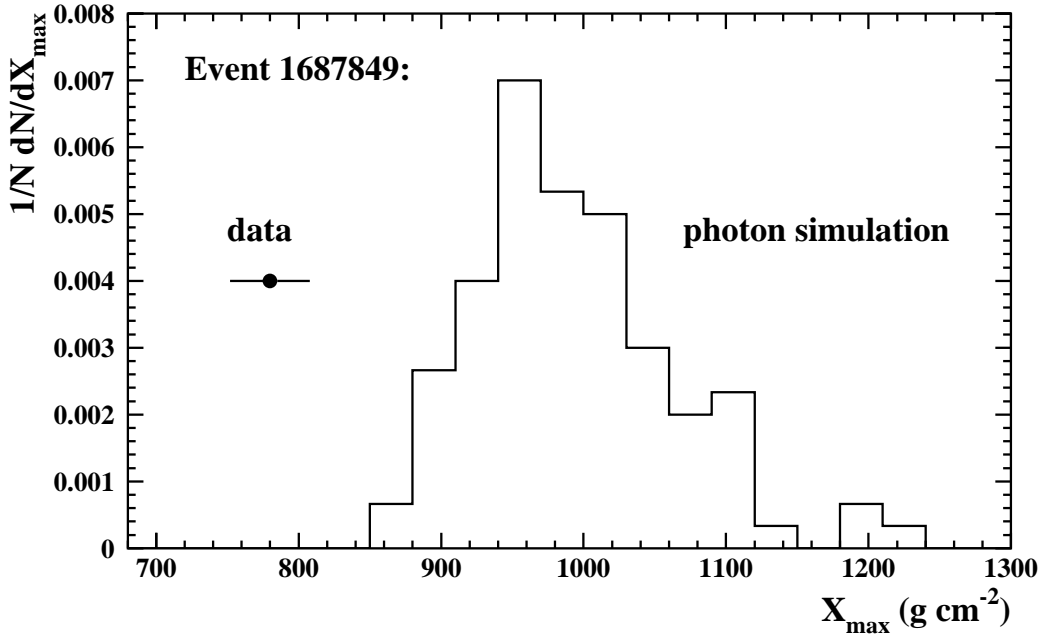}
\caption{
$X_{\rm max}$ measured in the shower shown in
Figure~\ref{fig-evprof} (point with error bar)
compared to the $X_{\rm max}^\gamma$ distribution
expected for photon showers (solid line).
}
\label{fig-evxmax}
\end{center}
\end{figure}

In Figure~\ref{fig-evxmax} the predictions for $X_{\rm max}^\gamma$
for a photon primary are compared with the measurement of 
$X_{\rm max} = 780$~g~cm$^{-2}$
for event 1687849 (Figure~\ref{fig-evprof}).
With $\langle X_{\rm max}^\gamma \rangle \simeq 1000$~g~cm$^{-2}$,
photon showers are 
on average expected to reach maximum at depths considerably greater than
that observed for real events.
Shower-to-shower fluctuations are large due to the LPM effect.
For this event, the expectation for a primary photon differs by
$\Delta_\gamma \simeq$ +2.9 standard deviations from the data,
where $\Delta_\gamma$ is calculated from
\begin{equation}
\label{eq1}
\Delta_\gamma = \frac{ <X_{\rm max}^\gamma> - X_{\rm max}}
{ \sqrt{ (\Delta X_{\rm max}^\gamma)^2 +
  (\Delta X_{\rm max}^{\rm stat})^2 } }~~.
\end{equation}

\begin{figure}[t]
\begin{center}
\includegraphics[height=8.7cm,angle=0]{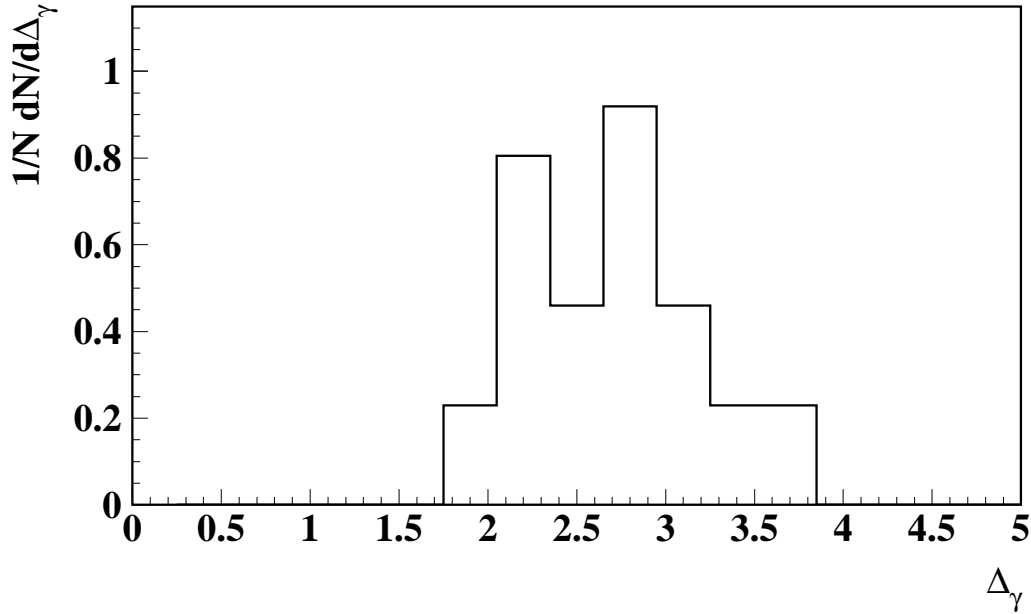}
\caption{
Distribution of differences $\Delta_\gamma$ in standard deviations
between primary photon prediction and data for the 29 selected events.
}
\label{fig-sigma}
\end{center}
\end{figure}

For all events, the observed $X_{\rm max}$ is well below the average
value expected for photons (see Table~\ref{tab-data}).
The differences $\Delta_\gamma$ between photon prediction and data
range from +2.0 to +3.8 standard deviations,
see Figure~\ref{fig-sigma} and Table~\ref{tab-data}.
It is extremely unlikely that all 29 events were initiated by photons
(probability $\ll$10$^{-10}$),
so an upper limit to the fraction of cosmic-ray photons above 10~EeV
can be reliably set.

Due to the limited event statistics, the upper limit cannot be
smaller than a certain value.
The relation between the minimum possible fraction
$f_\gamma^{\rm min}$ of photons that could be excluded for a
given number of events $n_{\rm m}$ (or:
the minimum number of events $n_{\rm m}^{\rm min}$ required to
possibly exclude a fraction $f_\gamma$) is given
by
\begin{equation}
\label{eq2}
f_\gamma^{\rm min}  = 1-(1-\alpha)^{1/n_{\rm m}}~,~{\rm and}~~~~
n_{\rm m}^{\rm min} = \frac{\ln(1-\alpha)}{\ln(1-f_\gamma)}~~,
\end{equation}
with $\alpha$ being the confidence level of rejection.
This holds for the case that no efficiency correction has to
be applied ($\epsilon_{\rm min} = 1$).
For 29 events and $\epsilon_{\rm min} \simeq 0.80$, the minimum
possible value for an upper limit
to be set at a 95\% confidence level is $\sim$~12\%.
The theoretical limit is reached only if
a photon origin is basically excluded for all events.

The calculation of the upper limit is based on the statistical
method introduced in~\cite{risse05} which is tailor-made for
relatively small event samples.
For each event, trial values $\chi^2 = \Delta_\gamma^2$ are calculated
with $\Delta_\gamma$ according to Eq.~(\ref{eq1}).
We distinguish between statistical and systematic uncertainties for
the depths of shower maximum.
The method in~\cite{risse05} is extended to allow for a correlated
shift of the observed $X_{\rm max}$ values for all selected events,
where the shifted value is drawn at random from a Gaussian distribution
with a width $\Delta X_{\rm max}^{\rm syst}$ = 23~g~cm$^{-2}$.
For the shifted data, new $\chi^2$ values are calculated from
Eq.~(\ref{eq1}).
Many such ``shifted'' event sets are generated from the data and 
compared to artificial data sets using photon simulations.
The chance probability $p(f_\gamma)$ is calculated to obtain artificial
data sets with $\chi^2$ values larger than observed as a function of the
hypothetical primary photon fraction $f_\gamma$.
Possible non-Gaussian shower fluctuations are accounted for in the
method, as the probability is constructed by a Monte Carlo technique.
The upper limit $f_\gamma^{\rm ul}$, at a confidence level $\alpha$, is
then obtained from 
$p (f_\gamma \ge \epsilon_{\rm min} f_\gamma^{\rm ul}) \le 1-\alpha$,
where the factor $\epsilon_{\rm min} = 0.80$ accounts for the different
detector acceptance for photon and nuclear primaries 
(Section~\ref{sec-data}).

\begin{figure}[t]
\begin{center}
\includegraphics[height=8.7cm,angle=0]{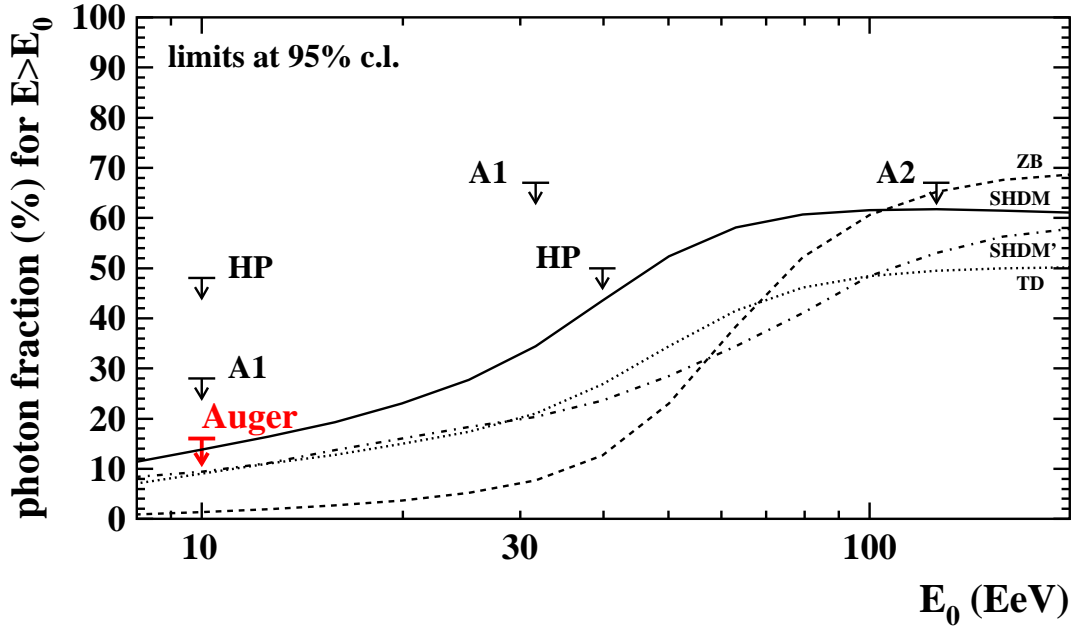}
\caption{
Upper limits (95\% c.l.) to the cosmic-ray photon fraction
derived in the present analysis (Auger) and obtained
previously from AGASA (A1)~\cite{shinozaki}, (A2)~\cite{risse05}
and Haverah Park (HP)~\cite{ave} data, compared to expectations
for non-acceleration models
(ZB, SHDM, TD from~\cite{models}, SHDM' from \cite{ellis}).
}
\label{fig-uplim}
\end{center}
\end{figure}

For the Auger data sample, an upper limit to the
photon fraction of 16\% at a confidence level of 95\% is derived.
In Figure~\ref{fig-uplim}, this upper limit is plotted together with
previous experimental limits and some illustrative estimates for
non-acceleration models. We have shown two different expectations for SHDM decay 
\cite{models,ellis} to illustrate the sensitivity to assumptions made about the
decay mode and the fragmentation, as well as the normalisation of the spectrum 
(see Figure~\ref{fig-specshdm}).
The derived limit is the first one based on observing the depth
of shower maximum with the fluorescence technique.
The result confirms and improves previous limits above 10~EeV
that came from surface arrays.
It is worth mentioning that this improved limit is achieved 
with only 29 events above 10~EeV,
as compared to about 50 events in the Haverah Park analysis and about 120 events
in the AGASA analysis.

%


\section{Discrimination power of surface array observables}
\label{sec-sd}

In the current analysis, data from the surface array are used
only to achieve a high precision
of reconstructed shower geometry in hybrid events. A single tank was
sufficient for this. However, observables registered by the surface
array are also sensitive to the primary particle type and can be
exploited for studies of primary photon showers.
In spite of the incomplete coverage of the array during the data taking 
period considered here (which means many events were poorly contained),
for about half of the selected events 
a standard array reconstruction~\cite{bertou05} can be performed.
Several observables can then be used for primary photon
discrimination, for instance the lateral distribution or the
curvature of the shower front~\cite{bertou00}.

An example for another observable is given by the {\it risetime}
of the shower signal in the detectors, one measure of the time
spread of particles in the shower disc.
For each triggered tank, we define a risetime
as the time for the integrated signal to go from 10\% to 50\% of its
total value.
By interpolation between risetimes recorded by the tanks at different
distances to the shower core, the risetime at 1000~m core distance
is extracted after correcting
for azimuthal asymmetries in the shower front.
The risetime is sensitive to the primary particle type because of its
correlation with shower muons and the depth of shower maximum:
contrary to the shower muons, electrons undergo
significant deflections with corresponding time delays.
Thus, larger values for the risetime are observed if the signal
at ground is dominated by the electromagnetic shower component.
Primary photon showers generally have fewer muons and, additionally,
the shower maximum is
closer to ground compared to showers from nuclear
primaries. Correspondingly, risetimes are expected to be
relatively large for photon primaries.

\begin{figure}[t]
\begin{center}
\includegraphics[height=8.7cm,angle=0]{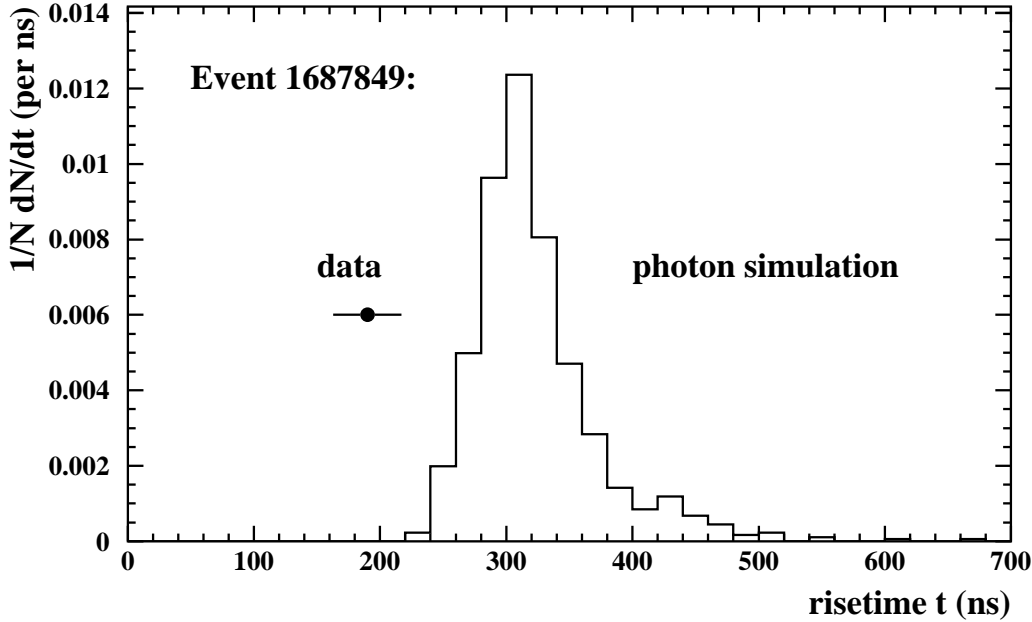}
\caption{
Example of risetime measured in an individual shower,
same as in Figure~\ref{fig-evprof}
(point with error bar) compared to the risetime distribution
expected for photon showers (solid line).
}
\label{fig-evrise}
\end{center}
\end{figure}

For the specific event shown in Figure~\ref{fig-evprof},
the measured risetime is compared
to the simulated distribution in Figure~\ref{fig-evrise}. 
For this and the other hybrid events with array reconstruction, the 
observed risetime does not agree well with the predictions
for primary photons, supporting the conclusion that a photon origin of
the observed events is not favored.
In future photon analyses, the independent information
on the primary particle from the Auger ground array and fluorescence
telescope data can be used to cross-check each other.
Combining the different shower observables will further
improve the discrimination power to photons.


\section{Outlook}
\label{sec-outlook}

The upper limit to the photon fraction above 10~EeV derived in this work
from the direct observation of the shower maximum confirms and
reduces previous limits from ground arrays.
The current analysis is limited mainly by the small number of events.
The number of hybrid events will considerably increase over the
next years, and much lower primary photon fractions can be tested.
Moreover, the larger statistics will allow us to increase the
threshold energy above 10~EeV
where even larger photon fractions are predicted by some models.

As an example, let us consider an increase in data statistics above 
10~EeV by about an order of magnitude compared to the current
analysis, as is expected to be reached in 2008/2009.
From Eq.~(\ref{eq2}), a sensitivity to photon fractions
down to $\sim$~1.5\% can be inferred.
More realistically, let us assume for the measured 
$X_{\rm max}$ values a distribution similar to the one currently
observed as in Figure~\ref{fig-sigma}. Then, an upper limit of $\sim$~5\%
could be achieved.
With the increased run time, 
a comparable number of events as for the present analysis would be reached
above 30--35~EeV.
If an upper limit similar to that reached here was found,
but at this higher energy, it
would be well below existing limits and severely constrain 
non-acceleration models.\footnote{A 36\% upper limit above 100~EeV has been
claimed recently from combining AGASA and Yakutsk data~\cite{troitsky};
however, the energies reconstructed for the AGASA events in that work
are in conflict with those given by the AGASA group.
}

The sensitivity of the hybrid analysis might be further improved in
the future by combining different shower observables measured in the same
event, such as depth of shower maximum, risetime and curvature.
We did not include ground array observables for the limit derived in 
this analysis since we wanted to independently check previous
ground array results.
Further information, e.g.~the width of the shower profile,
might also be added in future work to achieve better separation of
deeply penetrating nuclear primaries and primary photons. 

If only surface detector data is used and hybrid detection is
not required then the event statistics are increased by about an order
of magnitude.
Care must however be taken about a possible bias against photons in an array-only
analysis because of the different detector acceptance for photon and nuclear
primaries. Also, compared to the near-calorimetric energy determination
in the fluorescence technique, the energy estimated from array data
shows a stronger dependence on the primary type and is more strongly
affected by shower fluctuations.
Ways to reduce a possible photon bias and to place robust limits
to photons are being investigated.
For instance, the technique introduced in~\cite{ave} of comparing event
rates of near-vertical and inclined showers can be further exploited.


{\it Acknowledgements:}

We are very grateful to the following agencies and organisations for
financial support:
Gobierno de Mendoza, Comisi\'on Nacional de Energia At\'omica y 
Municipalidad de Malarg\"ue, Argentina;
the Australian Research Council;
Fundacao de Amparo a Pesquisa do Estado de Sao Paulo,
Conselho Nacional de Desenvolvimento Cientifico e Tecnologico and
Fundacao de Amparo a Pesquisa do Estado de Rio de Janeiro, Brasil;
National Science Foundation of China;
Ministry of Education of the Czech Republic (projects LA134 and
LN00A006);
Centre National de la Recherche Scientifique, Institut National de
Physique Nucl\'eaire et Physique des Particules (IN2P3/CNRS),
Institut National des Sciences de l'Univers (INSU/CNRS) et Conseil
R\'egional Ile de France, France;
German Ministry for Education and Research and Forschungszentrum
Karls\-ruhe, Germany;
Istituto Nazionale di Fisica Nucleare, Italy;
Consejo Nacional de Ciencia y Tecnologia, Mexico;
the Polish State Committee for Scientific Research (grant numbers
1P03D~01430, 2P03B~11024 and 2PO3D~01124), Poland;
Slovenian Research Agency;
Ministerio de Educaci\'{o}n y Ciencia (FPA2003-08733-C02, 2004-01198),
Xunta de Galicia (2003 PXIC20612PN, 2005 PXIC20604PN) and Feder Funds, Spain;
Particle Physics and Astronomy Research Council, UK;
the US Department of Energy, the US National Science Foundation, USA,
and UNESCO.

\appendix

\section{Distributions of quality cut variables}

\label{app-cut}

\begin{figure}[h]
\begin{center}
\includegraphics[height=6.5cm,angle=0]{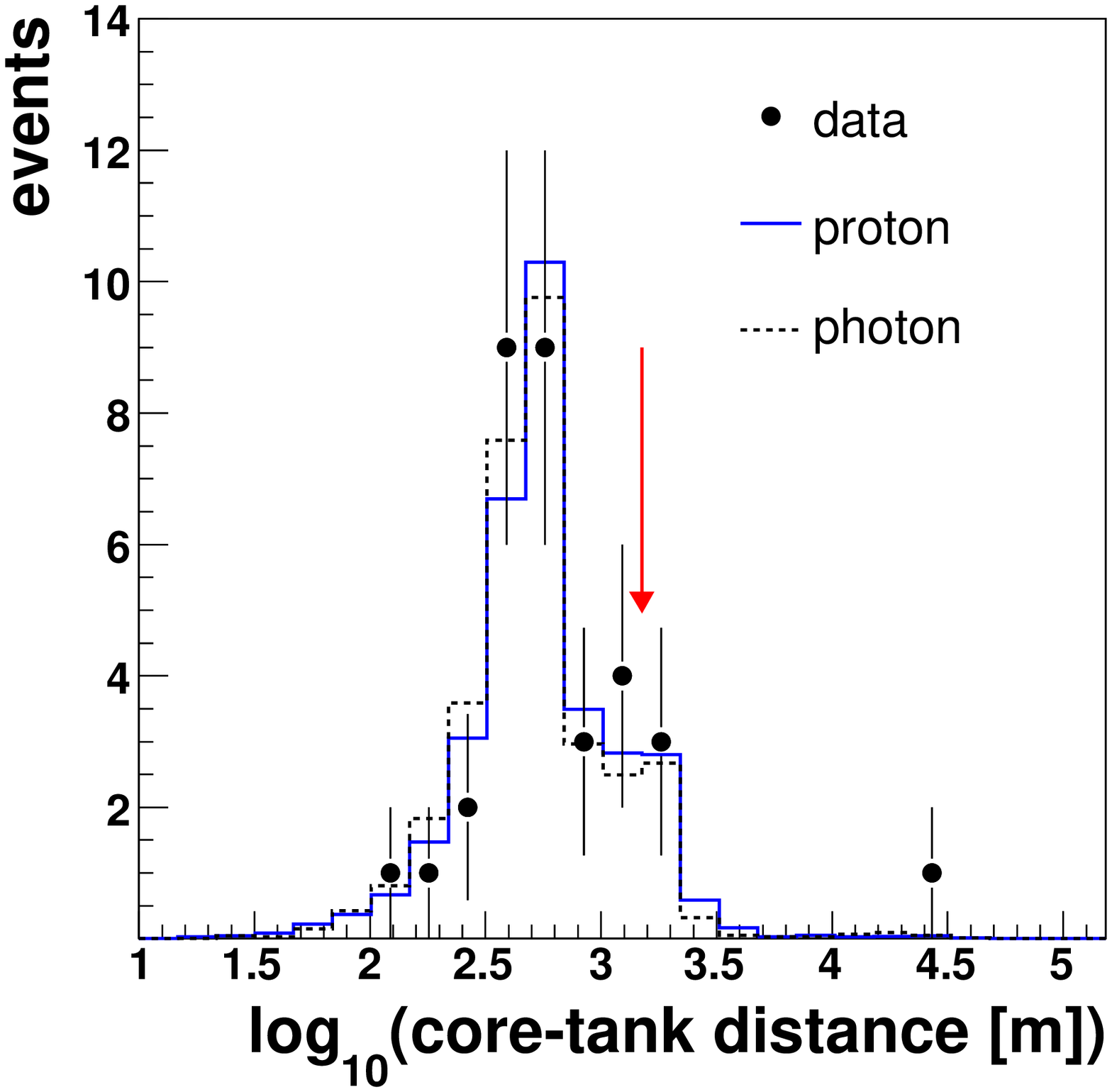}
\includegraphics[height=6.5cm,angle=0]{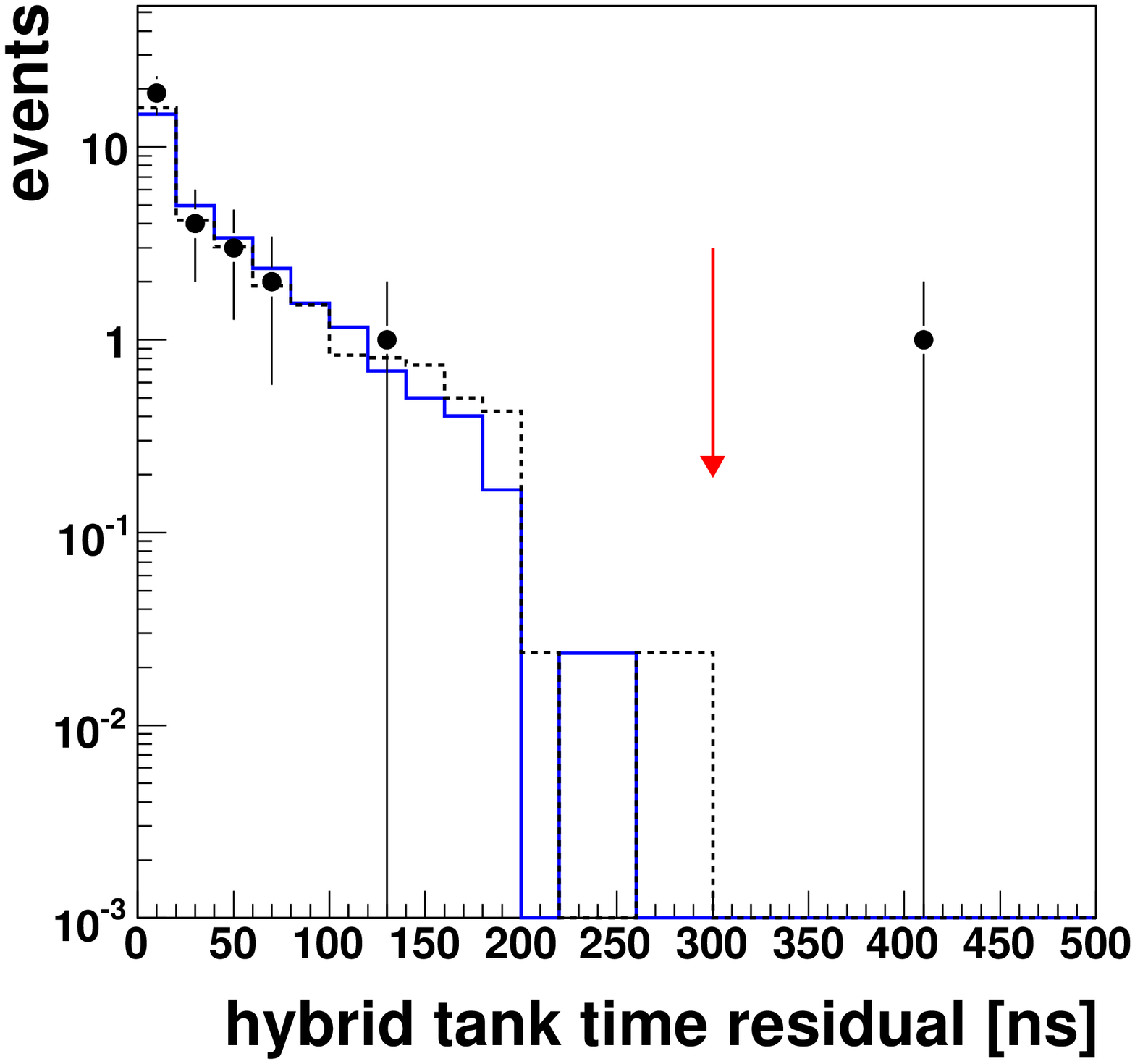}
\includegraphics[height=6.5cm,angle=0]{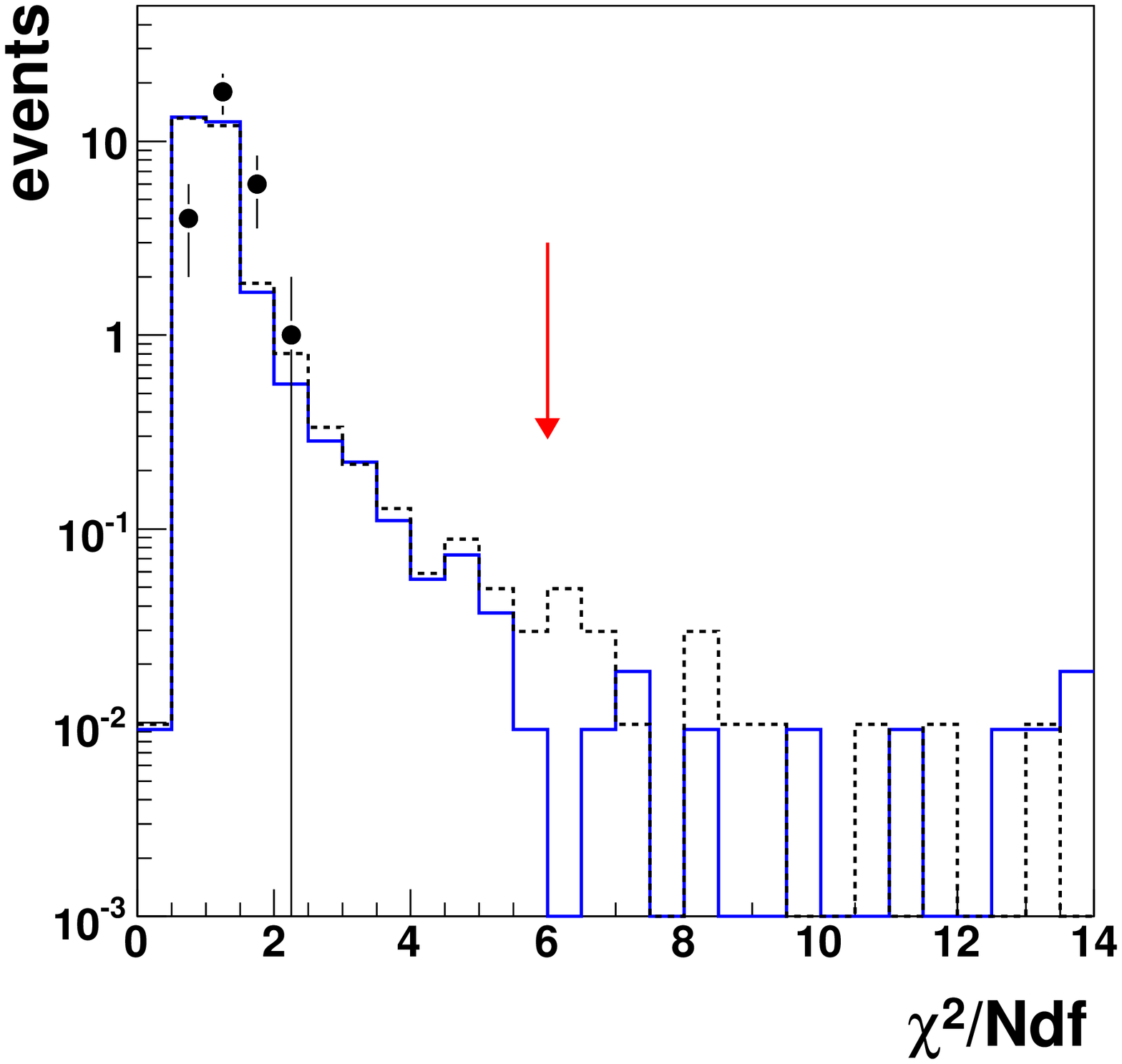}
\includegraphics[height=6.5cm,angle=0]{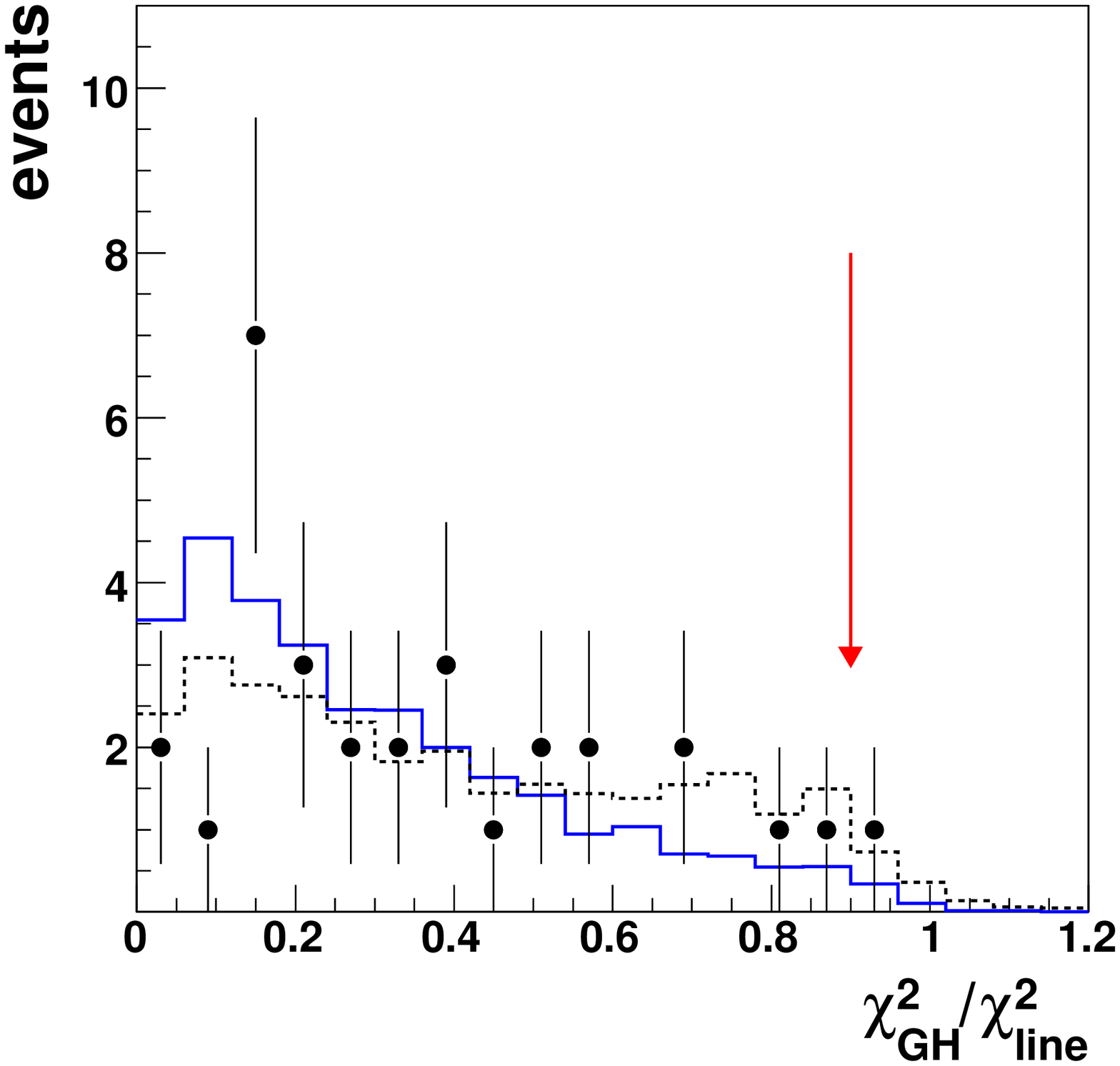}
\caption{
Distributions of variables after applying all quality cuts except
the one for the variable shown. The distributions are plotted for
data (filled circles), primary photons (dashed black histograms),
and primary protons (solid blue histograms).
The arrow indicates the cut position.
Plotted are distributions of distances of the tank with the
largest signal to the shower core (upper left panel), of the time
residual between that tank and the expected arrival time of the
shower front (upper right panel), of the reduced $\chi^2$ from the
Gaisser-Hillas profile fit (lower left panel), and of the ratio of this
reduced $\chi^2$ to that of a straight line fit (lower right panel).
}
\label{fig-appcut1}
\end{center}
\end{figure}

In Figure~\ref{fig-appcut1}, distributions of cut variables are plotted.
For each graph, all quality cuts (see Section~\ref{sec-data})
except the one for the variable shown were applied.

\section{Acceptance correction}

\label{app-acc}

The fraction of photons $f_\gamma$ in the cosmic-ray flux
integrated above an energy threshold $E_0$ is given by
\begin{equation}
f_\gamma(E\ge E_0) =
\frac{\int_{_{E_0}} \Phi_\gamma (E) dE}
{\int_{E_0} \Phi_\gamma (E) dE
+\sum_i\int_{E_0} \Phi_i (E) dE}
\end{equation}
where $\Phi_\gamma (E)$ denotes the differential flux of photons
and $\Phi_i (E),~ i={\rm p,He,...}$ the fluxes of nuclear primaries.

The fraction of photons $f_\gamma^{\rm det}$ as registered by
the detector is given by
\begin{equation}
f_\gamma^{\rm det}(E\ge E_0) =
\frac{
\int_{E_0} A_\gamma (E) \Phi_\gamma (E) dE 
}{
\int_{E_0} A_\gamma (E) \Phi_\gamma (E) dE
+\sum_i\int_{E_i} A_i (E) \Phi_i (E) dE
}
\end{equation}
with $A_\gamma (E)$ and $A_i (E)$ being the detector acceptances
to photons and nuclear primaries, respectively. $E_i$ denotes
the effective threshold energy for primary nucleus $i$.

Thus, the upper limit $f_\gamma^{\rm ul,det}$
obtained to the registered data,
$f_\gamma^{\rm ul,det} > f_\gamma^{\rm det}$, needs to be
corrected to resemble an upper limit to the fraction of
photons in the cosmic-ray flux. For the present analysis, a conservative
and model-independent correction is applied as follows.

$E_0$ corresponds to the analysis threshold energy assuming primary photons.
$E_i$ is related to $E_0$ by the ratios of the missing energy
corrections $m_\gamma$ (for photons) and $m_i$ (for nuclear primaries),
\begin{equation}
E_i = E_0 \cdot \frac{m_i}{m_\gamma}~.
\end{equation}
Since $m_\gamma \simeq 1.01$ and  $m_i\simeq 1.07-1.14$,
$E_i > E_0$. Thus, replacing $E_i$ by $E_0$,
\begin{displaymath}
f_\gamma^{\rm det}(E\ge E_0) >
\frac{
\int_{E_0} A_\gamma (E) \Phi_\gamma (E) dE
}{
\int_{E_0} A_\gamma (E) \Phi_\gamma (E) dE
+\sum_i\int_{E_0} A_i (E) \Phi_i (E) dE
}
\end{displaymath}
\begin{equation}
~~~=
\frac{
\int_{E_0} A_\gamma (E) \Phi_\gamma (E) dE 
}{
\int_{E_0} A_\gamma (E) \Phi_\gamma (E) dE
+\sum_i\int_{E_0} \frac{A_\gamma (E)}{\epsilon_i (E)} \Phi_i (E) dE
}~.
\end{equation}
In the last step, the acceptance ratio
$\epsilon_i (E) = A_\gamma (E) / A_i (E)$ was introduced.

From the fiducial volume cuts shown in Figure~\ref{fig-acc},
it can be seen that $A_\gamma \simeq {\rm const}$ in the energy
range of interest. Also, from Figure~\ref{fig-acc}
the minimum acceptance ratio $\epsilon_{\rm min} \le \epsilon_i (E) $
can be extracted
(in the current analysis, $\epsilon_{\rm min} = 0.80$).
Hence, it follows:

\begin{displaymath}
f_\gamma^{\rm det}(E\ge E_0) >
\frac{\int_{E_0} \Phi_\gamma (E) dE
}{
\int_{E_0}  \Phi_\gamma (E) dE
+ \frac{1}{\epsilon_{\rm min}} \sum_i\int_{E_0} \Phi_i (E) dE
}
\end{displaymath}
\begin{equation}
~~~>
\epsilon_{\rm min} \cdot
\frac{\int_{E_0} \Phi_\gamma (E) dE
}{
\int_{E_0}  \Phi_\gamma (E) dE
+ \sum_i\int_{E_0} \Phi_i (E) dE
}
= \epsilon_{\rm min} \cdot f_\gamma(E\ge E_0)~,
\end{equation}
where it was used that $\frac{1}{\epsilon_{\rm min}} > 1$.

Consequently, an upper limit $f_\gamma^{\rm ul}$ to the
fraction of photons in the cosmic-ray flux can conservatively
be calculated as
\begin{equation}
f_\gamma^{\rm ul} =
f_\gamma^{\rm ul,det} / \epsilon_{\rm min}
 > f_\gamma^{\rm det} / \epsilon_{\rm min} > f_\gamma ~.
\end{equation}

The upper limit obtained this way does not depend on 
assumptions for the differential fluxes $\Phi_\gamma (E)$
and $\Phi_i (E)$.

\end{document}